\documentclass[aps,pre,onecolumn,showpacs]{revtex4-1}

\usepackage{graphicx}

\begin{document}

\title{Fault heterogeneity and the connection between aftershocks and afterslip }
\author{
Eugenio Lippiello, 
Giuseppe Petrillo}
\affiliation{Department of Mathematics and Physics, University of Campania ``L. Vanvitelli'', Viale  Lincoln 5, 81100 Caserta, Italy}
\email{eugenio.lippiello@unicampania.it, giuseppe.petrillo@unicampania.it}
\author{
Fran\c{c}ois Landes}
\affiliation{iPhT, CEA Saclay, 91 Orsay, France;
Department of Physics and Astronomy, University of Pennsylvania, Philadelphia, PA; 
Department of Chemistry, Columbia University, New York, NY 10027, USA} 
\email{francoislandes@gmail.com}
\author{
Alberto Rosso}
\affiliation{LPTMS, CNRS, Univ. Paris-Sud, Universit\'e Paris-Saclay, 91405 Orsay, France}
\email{alberto.rosso@u-psud.fr}



\begin{abstract}
Whether aftershocks originate directly from the mainshock and surrounding stress environment or from afterslip dynamics is crucial to the understanding of the nature of aftershocks.
We build on a classical description of the fault and creeping regions as two blocks connected elastically, subject to different friction laws. 
We show analytically that, upon introduction of variability in the fault plane's static friction threshold, a non trivial stick-slip dynamics ensues.
In particular we support the hypothesis \citep{PA04} that the aftershock occurrence rate is proportional to the afterslip rate, up to a corrective factor that is also computed.
Thus, the Omori law originates from the afterslip's logarithmic evolution in the velocity-strengthening region.
We confirm these analytical results with numerical simulations, generating synthetic catalogs with statistical features in good agreement with instrumental catalogs. In particular we recover the Gutenberg-Richter law with a realistic $b$-value ($b\simeq 1$) when Coulomb stress thresholds obey a power law distribution.

\end{abstract}

\maketitle

\vskip0.3cm
\section{Introduction} 

A mainshock is followed by the increase of seismic activity caused by aftershocks as well as by
a significant time-dependent postseismic deformation known as afterslip.
The common interpretation is that afterslip is activated by
the stress increase due to the mainshock's coseismic instantaneous deformation
and mostly occurs in regions with a velocity strengthening rheology.
This interpretation dates back to the seminal paper by \citet{MSB91}
who analytically obtained  a hyperbolic time decay of the post-seismic deformation rate $\dot u(t) \sim 1/t$, for a velocity strengthening region .
The hyperbolic temporal decay is similar to the one usually observed for the aftershock occurrence rate, $\lambda(t)$:
\begin{equation}
\lambda(t)=\frac{K}{t+c},
\label{omori}
\end{equation}
and known as the Omori law.

The proportionality between  aftershock occurrence rate, $\lambda(t)$, and stress or strain rate, $\dot u(t)$, has been documented by the postseismic deformation measured after several large earthquakes 
\citep{PA04,PAR05,Hsu06,PA07,PA08,SY07,SL08,Sav10,CGHLLS18,PFMB18}.
\citet{PA04} explained the observed proportionality under the assumption that  
aftershocks are induced by afterslip. This result was supported by the analytical solution of a single spring-slider model under velocity strengthening friction which models brittle creeping. The proportionality  has been also used to obtain the friction parameters as well as the stressing rate, in a given velocity strengthening region, from the temporal behavior of the recorded aftershocks \citep{FPP17}.

\begin{figure}
\centering
\includegraphics[width=12cm]{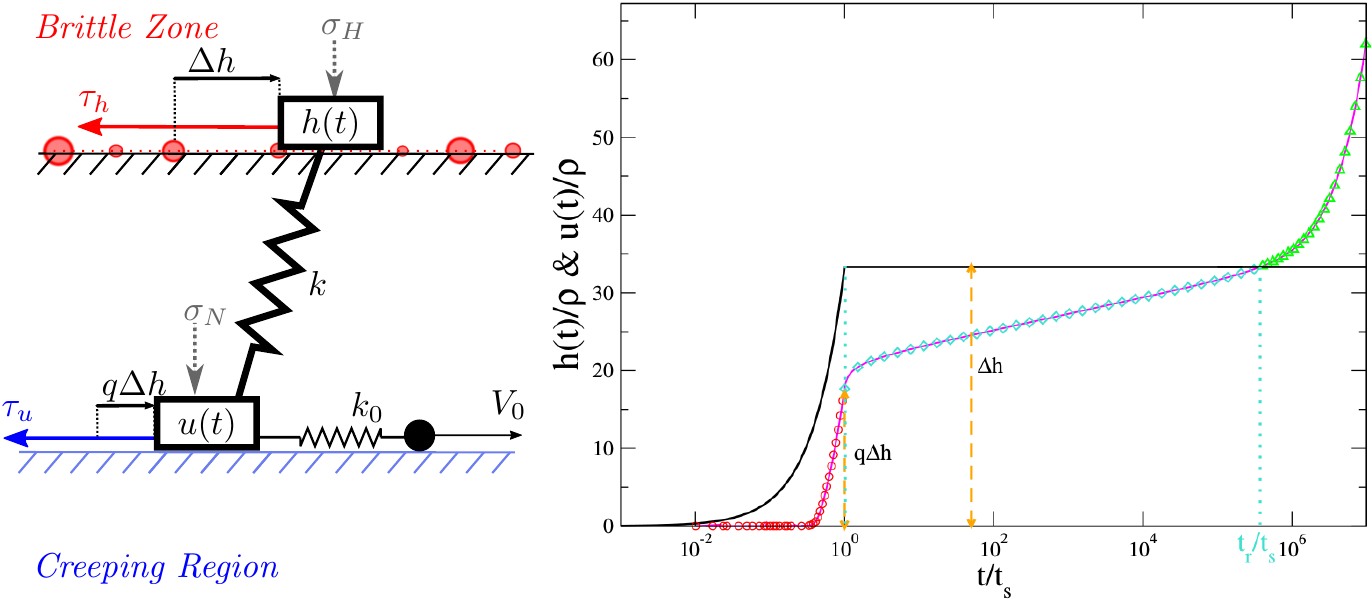}
\caption{
 (Left) The two block model. The block $H$, at the position $h(t)$, represents the fault which performs discrete jumps of fixed amplitude $\Delta h$ and can be stuck in different positions indicated by dots. The size of each dot represents the local value of $f^{th}$. The block $U$, at the position $u(t)$, is subject to a velocity strengthening friction  and is driven at a constant rate $k_0 V_0$. Both block are subject to confining pressures $\sigma_N$ and $\sigma_H$. 
 The exact solution of Eq.~(\ref{VS}) assuming the the slip starts at $t=0$ and ends and $t=t_s$ with the block $H$ moving at a constant velocity $V_S$. 
Distances are expressed in units of $\rho$ and time in unit of $t_s$. 
We use $\Delta h=33 \rho$, $V_0=10^{-5} \delta h/t_s$, and $k/k_0=33$.
The fault displacement $h(t)/\rho$ is plotted with a continuous black line whereas
the full analytical solution of Eq.~(\ref{VS}) for $u(t)/\rho $ is represented by a continuous grey line.
Different colors and symbols indicate the evolution of $u(t)$ in the three regimes: circles (slip regime), diamonds (afterslip regime) and triangles (interseismic regime).}
\label{picture}
 \label{hu_t}
\end{figure}

In this letter we present a minimal model of the fault as a sliding block connected to the  afterslip region treated, as in  \citet{MSB91,PA04},  as a second block with velocity strengthening rheology (see Fig.~\ref{picture}). 
In this two-block model the slip of the fault block induces the afterslip relaxation of the second block which in turns promotes further failures of the fault. 
The central assumption is that instability thresholds $f^{th}$, on the fault plane, is not uniform but random with a distribution $g(f^{th})$.
This  makes more realistic the description of the fault as a single-slider block.  
 Previous studies \citep{KL07,ALAA14}, indeed, have shown that the absence of heterogeneities is responsible for important differences between single-slider models and two-dimensional continuum models.
The presence of random thresholds allows us to analytically demonstrate that, without any assumption on the initial stress distribution, the proportionality $\lambda(t) \propto \dot u(t)$ is a stable feature of earthquake triggering.
We present analytical and numerical results of the model evolution.   

\vskip0.3cm
\section{The model}

We describe the fault as a block $H$ 
coupled by means of a spring of elastic coupling $k$ to a second block $U$ which represents the creeping region of the crust, with velocity strengthening friction.  
 \citet{MSB91}  identified the region where afterslip occurs in a zone of unconsolidated sediments above the fault whereas  \citet{PA04,PAR05,PA07} assumed the creeping region zone to be deeper because of the transition towards ductile behavior caused by the increase of temperature with depth. In other studies afterslip has been observed very close \citep{CAS99} or even on the fault within the seismogenic zone \citep{MSFK04, JBL06, Fre07}.
In our approach, the precise position of the velocity strengthening region is not relevant and we assume that the block $U$ is
embedded in a more extensive region creeping at a constant velocity $V_0$.
We model this interaction with a second spring of elastic constant $k_0$ whose free end moves with velocity $V_0$
(Fig.~\ref{picture}). To simplify the model we assume that both blocks move in the $V_0$ direction and, therefore, only scalar quantities can be considered.
Indicating with $u(t)$ and $h(t)$ the positions of the block $U$ and the block $H$ at the time $t$, respectively, the shear stress acting on $U$ is $k\left( h(t)-u(t)\right)+k_0\left( V_0 t -u(t)\right)$
and the friction force $\tau_u(t)$, under steady state condition, can be written as  
\begin{equation}
\tau_u(t)=\sigma_N \left(\mu_c + A \log \left (\frac{\dot u(t)}{V_c}\right) \right)
\end{equation}
where $\dot u(t)$ is the block velocity,  $\sigma_N$ is the effective normal stress, $\mu_c$ is the static friction coefficient when the block $U$ slides at the steady velocity $V_c$ and $A>0$ for a velocity strengthening material. 
In the overdamped limit, the constitutive equation for $U$ reads:
\begin{equation}
\tau_u(t)
= k\left( h(t)-u(t)\right)
+k_0\left( V_0 t -u(t)\right),
\label{VS}
\end{equation}
 which admits a stationary solution for $\dot{u}(t)=V_c=\frac{k_0}{k_0+k}V_0$.
In general one has 
\begin{equation}
\dot u(t)=V_c \exp \left (\frac{h(t)}{\rho}+\frac{t}{t_R}-\frac{\mu_c}{A}-\frac{u(t)}{\rho_0}\right)
\end{equation}
where $\rho_0 = \frac{A \sigma_N}{k + k_0}$ and $\rho = \frac{A \sigma_N}{k}$ are two characteristic length whereas
$t_R = \frac{A \sigma_N }{k_0 V_0}=\frac{\rho_0}{V_c}$ represents the long time scale associated with the extended creeping region at velocity $V_0$.


We next consider the evolution of the block $U$ starting from the 
stationary solution $\dot u(0)=V_c$ at the time $t=0$. 
 The solution for $u(t)$ was written in \citet{PA04} in term of the evolution of $h(t)$:
\begin{eqnarray}
u(t)&=&u(0)+\rho_0 \log \left( 1 + \frac  {1}{t_R}   F(t)  \right) \nonumber \\
F(t)&=&\int_0^t \exp \left(\frac {t'}{t_R} + \frac{(h(t')-h(0))}{\rho} \right) dt'.
\label{udit2}
\end{eqnarray}

Concerning the dynamics of the block $H$ 
we assume the Coulomb failure criterion (CFC): The instability is only controlled by the Coulomb stress $S_H(t)$ acting at time $t$ on the block $H$, with $S_H(t)=k(u(t)-h(t))-\mu_H \sigma_H(t)$. Here $k(u(t)-h(t))$ is the shear stress, $\mu_H$ is a friction coefficient and $\sigma_H(t)$ the effective normal stress represented by the normal stress reduced by the pore pressure. 
According to the CFC, under the assumption that $\sigma_H(t)$ is constant,  the block $H$ is unstable as soon as the shear stress overcomes a reference frictional stress $f^{th}$. 
Treating each slip as instantaneous implies that $H$ either is at rest or it slips in an infinitely short time.   
This approximation  corresponds to a vanishing nucleation size  coherently with the steady-state approximation in the constitutive equation (Eq.(\ref{VS})) \citep{RBZ96}.
Within our hypothesis, the dynamics of the two blocks can be written in terms of the Eqs.(\ref{udit2}) and split in three regimes:
\begin{itemize}
\item \textbf{Slip regime}. If at time $t=0$, $k(u(0)-h(0)) > f^{th}$, the position of the block $H$ becomes unstable and jumps of $\Delta h$ inducing the coslip of the block $U$ which depends on the precise dynamics of the block $H$. For example in Fig.~\ref{hu_t} we present an explicit solution when $h(t)$ moves steadily at velocity $V_s \gg V_0$ for a short time $t_s$. In all cases the block $U$ slips coseismically with the block $H$  
of an amount $q \Delta h$ with $q \in [0,1]$, so that the slip is given by:
\begin{eqnarray}
h(0) \rightarrow h(0) + \Delta h \\
u(0) \rightarrow u(0) +  q \Delta h.
\end{eqnarray}
The precise value of $q$ is a complicated function of model parameters and of the specific slip profile. In our minimal model we assume that $q$ is a fixed value equal for all slips. 

At the end of the slip, the block $H$ experiences a stress drop $k(1-q) \Delta h$, and if $k(u(t_s) - h(t_s))< f^{th}$ the fault is stuck again $h(t)=h(t_s)$.  
In the hypothesis of a uniform frictional stress $f^{th}$, the fault exhibits a trivial dynamics characterized by jumps of equal size ($\Delta h$) separated by a constant time interval. In this case the pre-stress value $f^{th}-k(u(0)-h(0))$ is peaked at a characteristic value and stress and seismic rate are not proportional.  Nevertheless, different sources of heterogeneity should affect the local value of the friction coefficient: The presence of asperities which induces fluctuations in the normal and pore pressure values, the variation of the basal friction coefficient or, again, the fluctuations of  $A$. These parameters are temperature and water dependent and are likely to vary along the fault. Thus it is reasonable that the frictional stress experienced after a slip $\Delta h$ is typically different from the previous one (see the variable intensity of pinning points -- red dots -- in  Fig.~\ref{picture}).
Concretely, after each slip it is reasonable that the block $H$ comes up against a new value of $f^{th}=f_1^{th}$ extracted from a distribution $g(f^{th})$ with 
a finite probability that $k(u(t_s)-h(t_s))>f_1^{th}$. Hence, the block is still unstable and accordingly can perform $n$ subsequent slips until the distance to failure  $\delta F_h=f_n^{th}-k\left(u(0)-h(0)\right)-n (1-q) \Delta h>0$. In this case, the slip instability corresponds to a single slip
of size $n\Delta h$ (a single earthquake) and the dynamics is characterized by earthquakes of different sizes. Large earthquakes can be, therefore, viewed as a succession of smaller seismic ruptures. 
After the slip the block $H$ is stuck in the novel position 
$h(t)=h(0)+n\Delta h$ and the  evolution of $u(t)$ is given by the explicit solution of Eqs.(\ref{udit2}):
\begin{eqnarray}
F(t)&=&t_R e^{\frac{n \Delta h}{\rho}} \left( e^{\frac{t}{t_R}}-1 \right), \\
u(t)&=&u(0)+n q \Delta h +\rho_0 \log \left( 1 + e^{\frac{n\Delta h}{\rho}}   \left( e^{\frac{t}{t_R}}-1   \right) \right).
\label{slipregime} 
\end{eqnarray}
Two distinct regimes can be identified.
\item \textbf{Afterslip regime}. At times shorter than $t_R$ one can replace $ t_R    \left( e^{\frac{t}{t_R}}-1   \right) \simeq t $ so that:
\begin{equation}
u(t) = u(0) + n q \Delta h +\rho_0 \log \left( 1 + \frac {e^{\frac{n\Delta h}{\rho}}}{t_R} t \right).
\label{afterslip1}
\end{equation}
In this regime the motion of $U$ increases the stress on $H$ inducing further slips of the block $H$, i.e. the aftershocks. This occurs if $ k \rho_0 \log \left( 1 + \frac{e^{\frac{n\Delta h}{\rho}}}{t_R} t_{AS} \right) =\delta F_h$, namely
\begin{equation}
t_{AS} = 
\frac{t_R}{e^{\frac{n\Delta h}{\rho}}}\left(e^{\frac{\delta F_h }{k\rho_0}}-1\right)
\quad \mbox{with} \quad t_{AS} \ll t_R. 
\label{tas}
\end{equation}  
\item \textbf{Interseismic regime}. At $t>t_R$ the motion of the block $U$ is dominated by the creeping velocity $V_0$. In this regime one can assume $  \left( e^{\frac{t}{t_R}}-1   \right) \simeq e^{\frac{t}{t_R}} $ and neglect the one in the logarithm of Eq.(\ref{afterslip1}). We obtain that the block $U$ slides at the steady velocity $ V_c=\frac{k_0}{k+k_0} V_0 $:
\begin{equation}
u(t)=V_c t + \mbox{const}.
\label{inters}
\end{equation}
\end{itemize}

\vskip0.3cm
\section{Statistics of slip events: analytical results}

In the following we explore the statistical features of the evolution 
of a single fault represented by the block $H$ which, because of the coupling with the block $U$, can experience multiple slip events.
The statistics of the single fault is expected to be representative of the statistics of  a population of faults with random initial stress conditions. 

The block $H$ exhibits a stick-slip dynamics with non trivial temporal correlations for a sufficiently broad distribution $g(f)$ which leads to a broad distributed distance to the next instability. In particular, we indicate with $t_0=0$ the time of the last slip and consider the probability $P_0(t)$ of no slip up to time $t$.
Since $u(t)$ is monotonically increasing with time (Eq.~(\ref{udit2})), $P_0(t)$ corresponds to the probability to extract a friction threshold $f_1^{th}$ larger than $k(u(t)-h(t))$, $P_0(t)=\int_{k(u(t)-h(t))}^{\infty}g(f)df$.
In the hypothesis that $g(f)$ does not present sharp discontinuities, the evolution of the block $H$ can be described as a time-dependent non-homogeneous Poisson process with $P_0(t)=\exp{\left(-\int_0^t \lambda(t')dt' \right)}$ where $\lambda(t)$ is the seismicity rate, i.e. the number of earthquake for unit time triggered by the mainshock. As a consequence, $\lambda(t)=-\frac{\partial \log \left(P_0(t) \right)}{\partial t}$ 
leading to 
\begin{equation}
\lambda(t)= Q\left(k(u(t)-h(t)) \right) k  \frac{\partial (u(t)-h(t))}{\partial t}
\label{lambdat1}
\end{equation}
with 
\begin{equation}
Q(x)= \frac{g\left(x\right)}{\int_{x}^{\infty}g(f)df}.
\end{equation}
Taking into account that $h(t)=const$, outside the very short slip regime, we find
\begin{equation}
\lambda(t)= Q\left(k(u(t)-h(t)) \right) k \dot{u(t)}. 
\label{lambdat}
\end{equation}
 Because of the presence of the time-dependent coefficient $Q\left(k(u(t)-h(t)) \right)$, Eq.(\ref{lambdat}) shows that in our model the seismic rate $\lambda(t)$ is not exactly proportional to the stress rate. However, we argue that in the afterslip regime $Q(k(u-h))$ can only depend logarithmically on time and the proportionality between aftershock and stress rate is, at first order, satisfied. More precisely,  we first focus on three types of distributions for the friction thresholds: a Gaussian distribution $g(f) = \sqrt{\frac{\alpha}{\pi}} \exp{\left(-\alpha \left(f-f_0 \right)^2 \right)}$ restricted to positive $f$, a power law distribution $g(f) = \frac{(\beta-1)f_0^{\beta-1}}{(f+f_0)^{\beta}}$ and an exponential distribution $g(f) = \frac{1}{\gamma} \exp\left(-\gamma f\right)$. In the case of the exponential distribution $Q(k(u-h))$  is exactly a constant whereas  it displays logarithmic time dependence for the power law ($Q(k(u-h)) \sim 1/(k(u-h)) \sim 1/\log(t)$) and for the Gaussian distribution ($Q(k(u-h)) \sim k(u-h) \sim \log(t)$). Therefore, using    $ \frac{\partial k(u-h)}{\partial t}=k \dot u(t) \sim 1/t$ in the afterslip regime  (Eq.~(\ref{afterslip1})),
we always obtain the  Omori hyperbolic decay $\lambda(t)\sim \dot u \sim 1/t $, with possible logarithmic corrections coming from $Q(t)$. An important exception to this behaviour is represented by the distributions with an upper cut-off $f_{\max}$. In this case, one can easily derive\footnote{for example by considering the uniform distribution $[0,f_{\max}]$} that $Q(k(u-h)) \propto 1/(f_{\max} - k(u(t)-h(t)))$ which can diverge at finite time.
Our results then shows that the key ingredients for the Omori law are represented by heterogeneities in the frictional stress values combined with the logarithmic stress relaxation induced by the velocity strengthening rheology. This behaviour is very general provided that the maximal values of the frictional stress are large compared to the increase of the stress that can be experienced during the afterslip phase.  

The presence of frictional heterogeneities also produces a non trivial slip size distribution
$p(n)$, which corresponds to the probability that the fault $H$ performs a slip of size $n \Delta h$. Assuming a small stress drop $k (1-q)\Delta h$ and the independence between subsequent jumps we can us the mapping to a record problem.
More precisely we neglect the stress change after each slip assuming $k(u(t)-h(t))-k(1-q)\Delta h \simeq k(u(t_0)-h(t_0))$, and calculate $p(n)$  as the probability to draw $n+1$ independent and identically distributed random variables $f_0^{th},f_1^{th},...,f_n^{th}$, such that $f_j^{th}< f_0^{th}$ for $j \in [1,n-1]$ and $f_n^{th}> f_0^{th}$. 
 In this case we find $p(n) \propto 1/\left(n^2+n\right)$ 
for any probability density function $g(f_0)$ and  independently of its domain $(f_{\min}, f_{\max})$ (here we assumed $f_{\min}\geq0$ and $f_{\max}=\infty$). To show it, we remark: 
 $ p(n)= \int _{f_{\min}}^{f_{\max}} df_0 g(f_0) \left(P_<(f_0)\right)^{n-1} \times P_>(f_0)$, where $P_<(f_0) = \int _{f_{\min}}^{f_0} df' g(f')$ is the probability to draw a threshold smaller than $f_0$ and, similarly, $P_>(f_0) = 1-P_<(f_0)$ which gives $g(f_0)=\frac{d P_<(f_0)}{df_0}$.
As a consequence, $ p(n)= \int _{f_{\min}}^{f_{\max}} df_0 \frac{d P_<(f_0))}{df}$ $\left (\left(P_<(f_0)\right)^{n-1}-\left(P_<(f_0)\right)^{n} \right)$ $= \frac{1}{n}-\frac{1}{n+1}$. For sufficiently large $n$ we obtain the power law behavior $p(n)\sim n^{-\eta}$ with $\eta=2$. This is in qualitative agreement with the Gutenberg-Richter (GR) law for the magnitude distribution. Indeed, taking into account that  
the size of a slip $n\Delta h$ is proportional to the seismic moment  released during a slip instability, the GR law 
combined with the logarithmic relation between magnitude and seismic moment corresponds to a power law behavior $p(n)\sim n^{-\eta}$ with $\eta=1+(2/3)b$, where $b\simeq 1$ is the coefficient of the GR law.

\vskip0.3cm
\section{Numerical Results}

In numerical simulations, the block $H$ is at rest at time $t_0$. We consider $k(u(t_0)-h(t_0))<f^{th}$ and $u(t)$ evolving according to Eq.(\ref{inters}) which corresponds to the interseismic regime. 
The evolution of the block $U$ increases the shear stress  $\tau_h(t)$ and the first slip occurs at the time $t_0+t_M$ with
\begin{equation}
t_{M}=\frac{1}{k V_c} \delta F_h =t_R \frac{\delta F_h}{k \rho_0}
\label{tm1}
\end{equation}
obtained by the inversion of Eq.(\ref{inters}). The block $H$, then performs $n$ subsequent slips before reaching the stable condition $\tau_h^{(n)}<f_n^{th}$ which corresponds to an earthquake of size $n\Delta h$ occurred at the time $t_0+t_M$. 
The subsequent slip will occur at a time $t_0+t_M+\delta t$ where $\delta t$ is the minimum between 
$t_{AS}$ and $t_M$ given by Eq.(\ref{tas}) and Eq.(\ref{tm1}), respectively. If $\delta t=t_{AS}$ the new slip is correlated to the first one and they are considered belonging to the same seismic sequence. The process is iterated and therefore the sequence can contain many correlated earthquakes. A new sequence starts as soon as $\delta t=t_M$. In numerical simulations we fix the value $\Delta h=10\rho$ in order to have a clear time separation between $t_{AS}$ and $t_M$ (Eq.s(\ref{tas}),(\ref{tm1})) and  
vary the only the $g$-related parameters $\alpha,\beta,\gamma,f_0$.

\begin{figure}
\includegraphics[width=12cm]{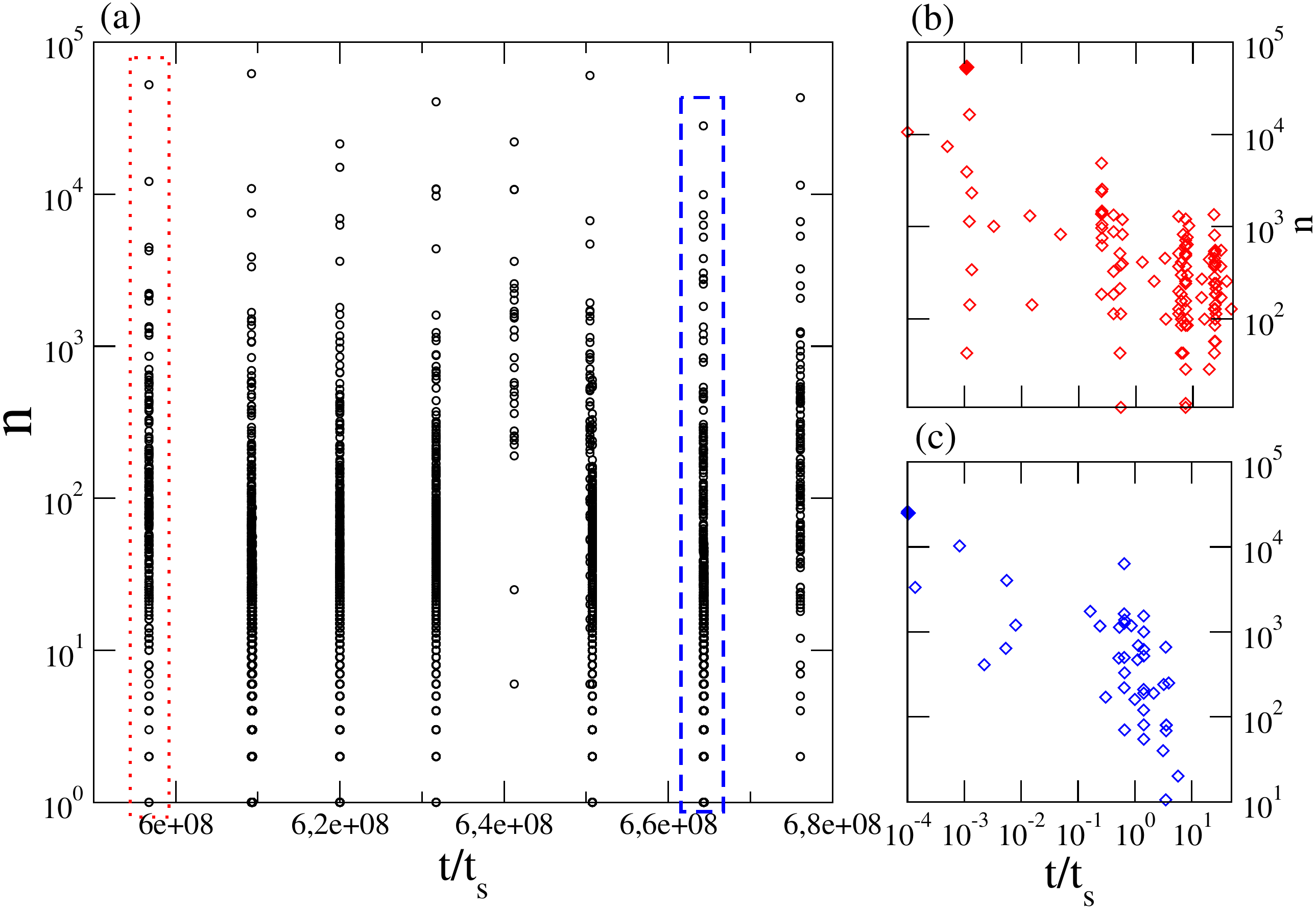}
\caption{(Panel a) A typical numerical catalog with a power law distributed $g(f)$ with , $\beta=5$, $f_0=10$ and $\Delta h/\rho=10$. (Panel b)
Zoom inside the temporal regions inside the dotted rectangle to show a main-aftershock sequences with two foreshocks. Time has been shifted to have the first event of the sequence at the time $t=10^{-4}t_s$. The mainshock is indicated by a filled symbol.   (Panel c) As for panel b for the region inside the dashed rectangle, corresponding to a sequence  without foreshocks.}
 \label{slip}
\end{figure}

A typical numerical sequence is plotted in Fig.~\ref{slip}a, where each point corresponds to an event with occurrence time $t$ and size $n \Delta h$.
The presence of correlated sequences, i.e. the clusters, is manifest.
In the majority of cases (more than $70\%$) the first event of the cluster is also the largest event belonging to the same sequence (Fig.~\ref{slip}c). In the remaining sequences the largest event, the mainshock, is anticipated by few smaller events, the foreshocks (Fig.~\ref{slip}b).

\begin{figure}
\includegraphics[width=12cm]{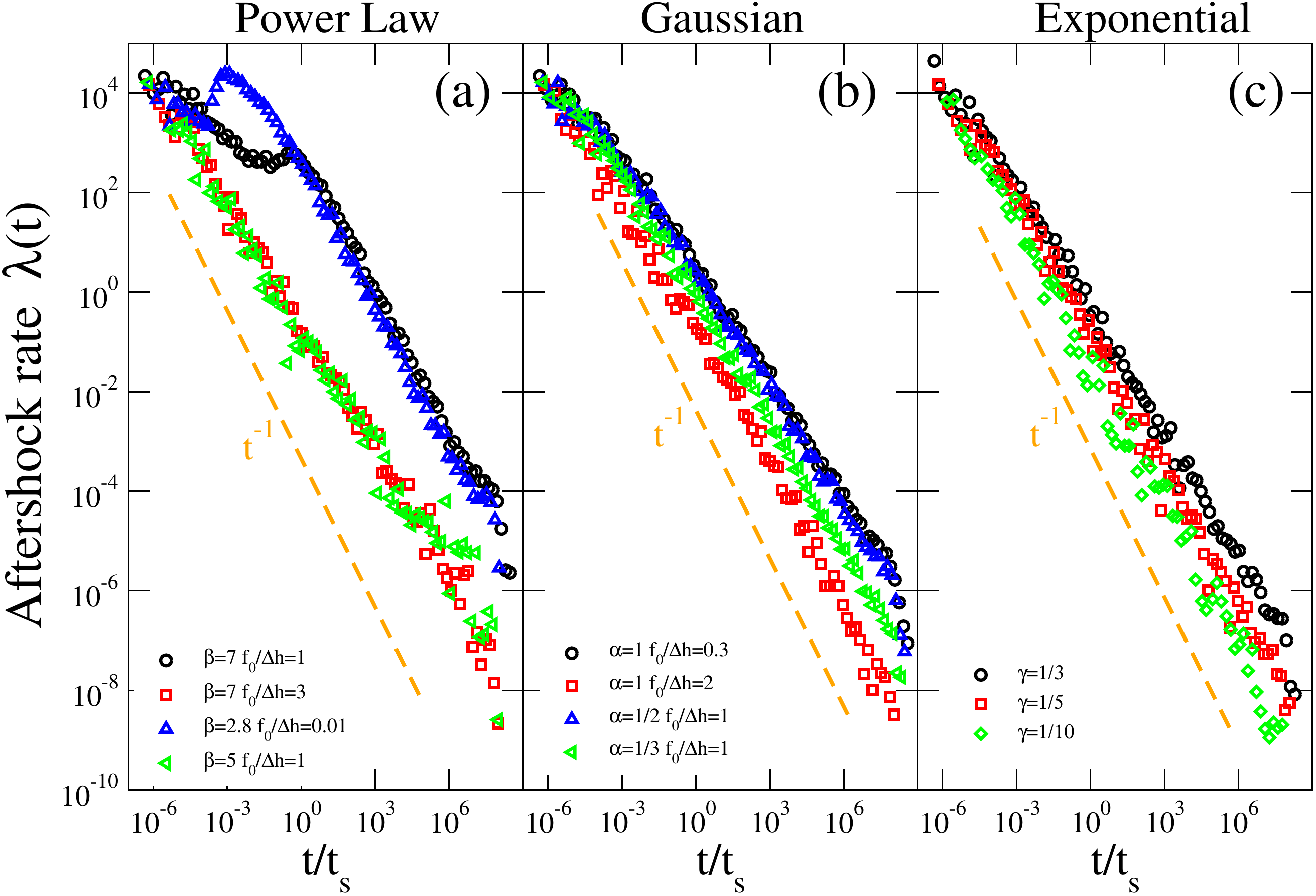}
\caption{The aftershock rate $\lambda(t)$ as a function of $t/t_s$, with $t$ the time since the mainshock, for three types of distributions: Power law (panel a), Gaussian (panel b) and exponential (panel c). Different symbols correspond to different parameters as in the figure legend. The dashed lines  represent the Omori decay $\lambda(t) \sim 1/t$.}
 \label{pt}
\end{figure}

In Eq.(\ref{tas}) we define aftershocks as those events following the first event of the sequence. In the following we adopt the more standard definition as those events following the mainshock. Since the number of foreshocks is always much smaller than the aftershock one, statistical features of aftershocks do not depend on the specific definition.
Fig.~\ref{pt} shows the rate of aftershocks $\lambda(t)$  as a function of time since the mainshock  which is in good agreement with the hyperbolic time decay predicted by the Omori law, for any distribution $g(f)$ and for different choices of parameters. 

\begin{figure}
\includegraphics[width=12cm]{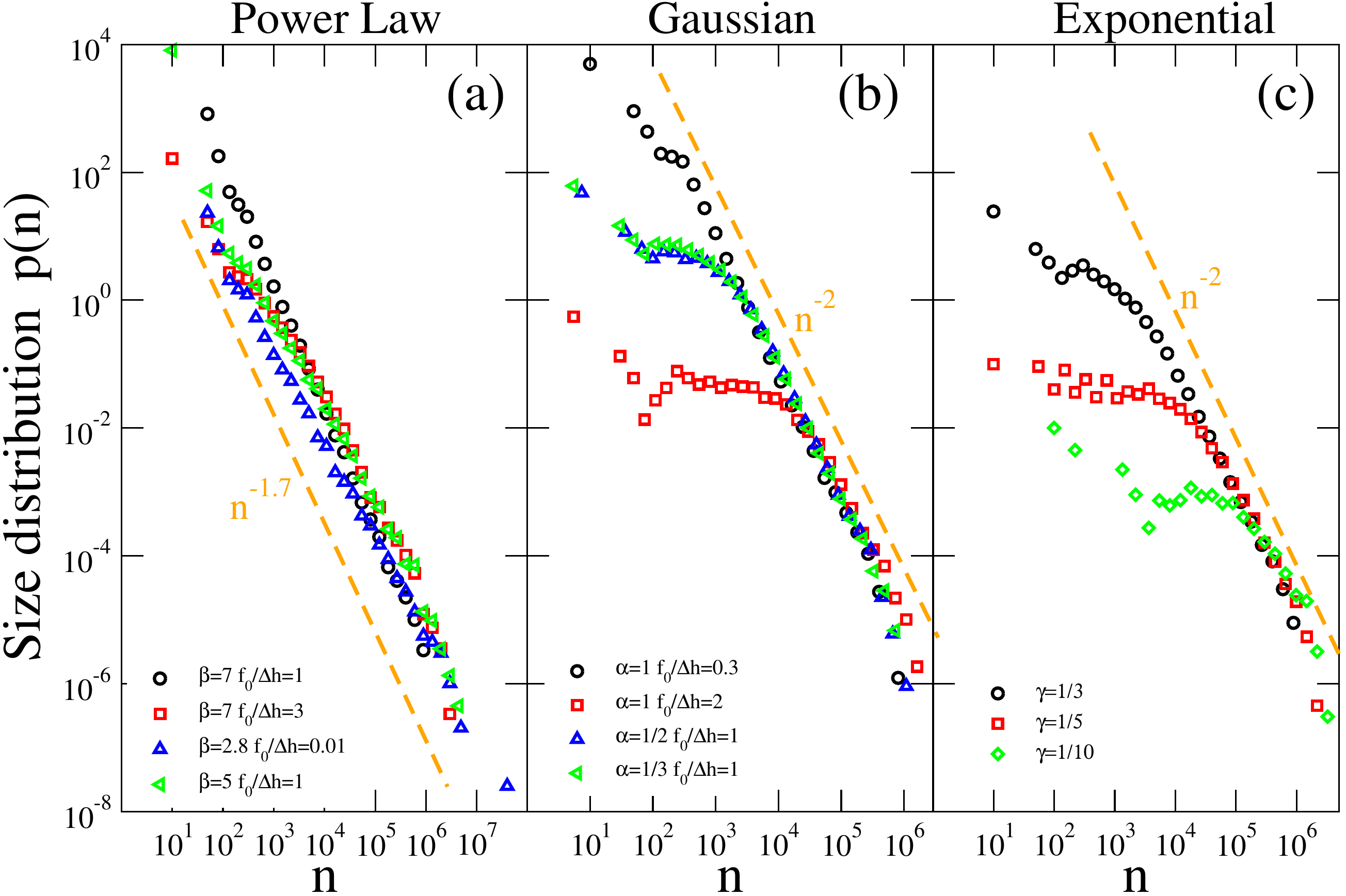}
\caption{
The size distribution $p(n)$ for three types of distributions:
Power law (panel a), Gaussian (panel b) and exponential (panel c).
Different symbols correspond to different parameters as in the figure legend. The dashed lines correspond to the best-fit power law decay $p(n) \sim n^{-\tau}$. We find $\tau \in (1.6,1.7)$ for the power law $g(f)$, and $\tau \simeq 2$ for the two other choices of the distribution.}
 \label{pn}
\end{figure}

In Fig.~\ref{pn} we plot the slip size distribution $p(n)$ in the synthetic catalog.
Results of numerical simulations (Fig.~\ref{pn}) show that for all choices of the distribution $g(f)$ and for different values of parameters, $p(n)$ exhibits a power law decay. 
We find the exponent $\eta \simeq 2$ for the Gaussian and exponential $g(f)$. Very interestingly, for a power law distributed $g(f)$  and for a wide range of $\beta$-values ($\beta \in [3,8]$),  we find  $\eta \in [1.6,1.7]$ corresponding to a $b$-value close to $1$ which is in quantitative agreement with the GR law of instrumental catalogs.

\vskip0.3cm
\section{Conclusions}

We have presented a minimal model for a seismic fault described as
a system of two elastically interconnected blocks, with a block representing the fault plane being subject to a random frictional force as opposed to another block having a velocity strengthening rheology. 
The evolution of the model presents the typical stick-slip behavior of real fault systems with mainshocks followed by aftershocks distributed in time according to the Omori law. Furthermore, the model reproduces the GR with a realistic $b$-value when the friction distribution $g(f)$ is a power law. This supports the hypothesis of power law distributed Coulomb stress thresholds in agreement with other indications of the self-similar nature of seismic occurrence
\citep{Sch02}.

The main difference with previous results is the presence 
of  randomness in the frictional thresholds. This  makes the distance to the failure  in our model always broadly distributed and allows us to avoid any
assumption on the pre-stress distribution. 
In our study, indeed,  the stress conditions before each slip instability originate from the previous stage of earthquake occurrence.
This is a novel result with respect to other scenarios where
 the Omori decay  is  obtained only starting from a population of fault patches in which pre-stresses are uniformly distributed. 
Assuming the CFC, for instance, 
the number of faults which slip in the time interval $\Delta t$ are those with $k(u(0)-h(0)) \in [f^{th}-\dot \tau(0) \Delta t,f^{th}]$ and are proportional to the stress rate $\dot \tau(0)$ for a uniformly distributed   $f^{th}-k(u(0)-h(0))$.  In particular, under stationary conditions $\dot \tau(t)=const$ this hypothesis gives a steady seismicity rate, whereas the Omori decay is obtained when the stress on the fault increases because of afterslip relaxation \citep{PAR05,PA07,PFMB18}. In an alternative interpretation \citep{Die94}, still assuming that pre-stresses are uniformly distributed, aftershocks are produced by the response of a population of rate weakening patches to a coseismic stress change. In presence of a rate weakening friction, indeed,  the relationship between $\lambda(t)$ and $\dot u(t)$ is no further linear \citep{HS06} and $\lambda(t)$ decays consistently with the Omori law in response to a stress drop.
\citet{Sav10} has shown that the combination of this relationship,
which holds for a velocity weakening fault \citep{Die94}, with the hypothesis that the driving stress is  generated by afterslip evolution \citep{PA04} leads to an improved description of post-seismic relaxation.
A rate weakening description of the fault block $H$ can be incorporated in our two-block model instead of the more simple CFC. 
In our approach we do not include  
a rate weakening description of the block $H$ and keep the more simple CFC since it allows us to perform analytical calculations and, at the same time, makes numerical simulations much more simple.
Nevertheless, because of heterogeneities in the frictional thresholds and since slip instabilities occur on the instantaneous time scale $t_s$, details of the friction law acting on the $H$-block are not expected to be relevant. The main effects of the friction laws should reflect  in changes of the fault slip $\Delta h$ which, however, is not a relevant parameter of our model. Indeed no significant differences are observed in numerical simulations where $\Delta h$ is not constant but is randomly extracted from a Gaussian distribution.

The comparison with instrumental aftershock sequences would be the subsequent step to support our main conclusions. A further aspect, not investigated in this paper, concerns the understanding of the  mechanism leading to the presence of small earthquakes, i.e. the foreshocks, which  in our numerical simulations anticipate the occurrence of large earthquakes. This could provide new insights on the outstanding question of the nature of foreshocks \citep{dAGGL16,LGMGdA17}.
As a further step, the two-block model can be  natural generalized to the many-block Burridge-Knopoff (BK) model \citep{BK67} coupled with a velocity strengthening region.
Recent studies \citep{Jag10,JK10,JLR14,LGGMD15,dAGGL16,LL16} have shown that, introducing an intermediate time scale for relaxation in a cellular automata version of the BK model, one recovers statistical features of instrumental catalogs, such as the Omori law and the GR law with a realistic $b$-value.
 Our results represent a justification for this class of models and provide insights in the mechanisms leading to realistic aftershock features.

\vskip 0.5cm
{\bf Data and Resources} No data were used in this paper.

\vskip 0.5cm
{\bf Acknowledgments}
We thank Eduardo Jagla and Hugo Perfettini for useful discussions. 
This work was partially supported by the grant from the Simons Foundation, for FL (\# 454935 Giulio Biroli, \# 327939 Andrea Liu, \# 454951 David Reichman)

\end{document}